\documentclass[pra,twocolumn,preprintnumbers,amsmath,amssymb,superscriptaddress]{revtex4-1}
\usepackage[paperwidth=210mm,paperheight=297mm,centering,hmargin=2.0cm,vmargin=2.6cm]{geometry}
\usepackage{graphicx}% Include figure files
\usepackage{dcolumn}% Align table columns on decimal point
\usepackage{bm}% bold math
\usepackage{upgreek} 
\usepackage{amsmath}
\usepackage{url}
\usepackage{hyperref}
\usepackage{lipsum}

\usepackage{dsfont}
\usepackage{mathrsfs}
\usepackage{bigints}
\usepackage{bbold}
\usepackage{amsthm}
\usepackage{setspace}
\usepackage{epstopdf}
\usepackage{color}

\newcommand{\beq}{\begin{equation}}
\newcommand{\eeq}{\end{equation}}
\newcommand{\bea}{\begin{align}}
\newcommand{\eea}{\end{align}}
\newcommand{\beqa}{\begin{eqnarray}}
\newcommand{\eeqa}{\end{eqnarray}}
\newcommand{\e}{\mathrm{e}}
\newcommand{\w}{\omega}
\newcommand{\ket}[1]{\left| #1 \right\rangle}
\newcommand{\bra}[1]{\left\langle #1 \right|}

\newcommand{\ketbra}[2]{\left|#1\right\rangle\hskip-1mm\left\langle #2\right|}

\newcommand{\vk}{\bm{k}}
\newcommand{\vecr}{\bm{r}}
\newcommand{\tr}{\operatorname{tr}}

\begin{document}
\title{Intrinsic and environmental effects on the interference properties of a high-performance quantum dot single photon source}

\author{Stefan Gerhardt}
\affiliation{Technische Physik and Wilhelm Conrad R\"ontgen Research Center for Complex Material Systems, Physikalisches Institut,
Universit\"at W\"urzburg, Am Hubland, D-97074 W\"urzburg, Germany}
\author{Jake Iles-Smith}
\affiliation{Department of Photonics Engineering, Technical University of Denmark, \O rsteds Plads, 2800 Kgs. Lyngby}
\author{Dara P. S. McCutcheon}
\affiliation{Quantum Engineering Technology Labs, H. H. Wills Physics Laboratory and Department of Electrical and Electronic Engineering, 
University of Bristol, BS8 1FD, UK}
\author{Yu-Ming He}
\affiliation{Technische Physik and Wilhelm Conrad R\"ontgen Research Center for Complex Material Systems, Physikalisches Institut,
Universit\"at W\"urzburg, Am Hubland, D-97074 W\"urzburg, Germany}
\affiliation{Hefei National Laboratory for Physical Sciences at the Microscale and Department of Modern Physics, $\&$ CAS Center for Excellence and Synergetic Innovation Center in Quantum Information and Quantum Physics, University of Science and Technology of China, Hefei, Anhui 230026, China}
\author{Sebastian Unsleber}
\affiliation{Technische Physik and Wilhelm Conrad R\"ontgen Research Center for Complex Material Systems, Physikalisches Institut,
Universit\"at W\"urzburg, Am Hubland, D-97074 W\"urzburg, Germany}
\author{Simon Betzold}
\affiliation{Technische Physik and Wilhelm Conrad R\"ontgen Research Center for Complex Material Systems, Physikalisches Institut,
Universit\"at W\"urzburg, Am Hubland, D-97074 W\"urzburg, Germany}
\author{Niels Gregersen}
\affiliation{Department of Photonics Engineering, Technical University of Denmark, \O rsteds Plads, 2800 Kgs. Lyngby}
\author{Jesper M\o rk}
\affiliation{Department of Photonics Engineering, Technical University of Denmark, \O rsteds Plads, 2800 Kgs. Lyngby}
\author{Sven H\"ofling}
\affiliation{Technische Physik and Wilhelm Conrad R\"ontgen Research Center for Complex Material Systems, Physikalisches Institut,
Universit\"at W\"urzburg, Am Hubland, D-97074 W\"urzburg, Germany}
\affiliation{SUPA, School of Physics and Astronomy, University of St Andrews, St Andrews, KY16 9SS, United Kingdom}
\author{Christian Schneider}
\affiliation{Technische Physik and Wilhelm Conrad R\"ontgen Research Center for Complex Material Systems, Physikalisches Institut,
Universit\"at W\"urzburg, Am Hubland, D-97074 W\"urzburg, Germany}

\date{\today}

\begin{abstract}

We report a joint experimental and theoretical study of the interference properties of a single photon source based on a 
In(Ga)As quantum dot embedded in a quasi-planar GaAs-microcavity. Using resonant laser excitation with a 
pulse separation of 2 ns, we find near-perfect interference of the emitted photons, and a corresponding indistinguishability of 
$\mathcal{I} = (99.6\,^{+\,0.4}_{-\,1.4})\%$. For larger pulse separations, quasi-resonant excitation conditions, 
increasing pump power or with increasing temperature, the interference contrast is progressively and notably reduced. 
We present a systematic study of the relevant dephasing mechanisms, 
and explain our results in the framework of a microscopic model of our system. 
For strictly resonant excitation, we show that photon indistinguishability is independent of pump power, but strongly 
influenced by virtual phonon assisted processes which are not evident in excitonic Rabi oscillations. 
\end{abstract}
\maketitle

\section{Introduction}

A central requirement for the implementation of single photons in quantum communication, quantum networks, linear optical quantum 
computing and quantum teleportation is their degree of indistinguishability~\cite{Pan2012, Kok2007, Brien2007, Nilsson2013, Dietrich2016}. 
To this end, cold atoms, single ions, isolated molecules, optically active defects in diamonds, silicon carbide, and layered materials, 
have all been identified as competitive sources of non-classical 
light~\cite{McKeever2004, Diedrich1987, Basche1992, Brouri2000, Kurtsiefer2000, Castelletto2014, He2015, Tonndorf2015}. 
In the solid state, to date, epitaxially grown self-assembled semiconductor quantum dots (QDs) have proven the most 
promising candidates for producing single photons with high single-photon purity 
combined with high quantum efficiency~\cite{Aharonovich2016,Michler2000,Santori2002,lodahl2015interfacing,gazzano13}. 
More recently, by embedding QDs in optical microcavities, the possibility to also generate highly indistinguishable photons has 
emerged~\cite{He2013,Ding2016,somaschi2016near,He2016,Iles-Smith2017_2}.

The indistinguishability condition can only be met if the overlap of the single-photon wavepackets in frequency, polarization, space and time is sufficient. 
Ideally a QD produces photons with Fourier limited wave packets, such that their temporal extension can be 
expressed as $T_2 = 2\,T_1$, where $T_1$ is the exciton lifetime. In reality, since a QD is embedded 
in a typically non-ideal host medium, one must combat dephasing channels such as phonon-coupling, or spectral wandering induced by coupling to carriers. 
If this coupling acts on the QD exciton on a timescale smaller, or comparable to photon emission events, the spectral width of the emitted 
wavepackets is broadened according to $\frac{1}{T_2} = \frac{1}{2T_1}$+$\gamma$, with the characteristic dephasing time $T^*_2$ = 1/$\gamma$~\cite{Kaer2013}. 
Furthermore, if the timing of emission events has some stochastic uncertainty, this will also reduce the expected 
wavepacket overlap and lead to a reduced indistinguishability~\cite{Kaer2013,uns15}. 

Previous experimental and theoretical studies have identified that the dominant dephasing processes in QD systems, namely spectral wandering and 
electron-phonon scattering, occur on two different timescales~\cite{Thoma2016,PhysRevB.87.081308,uns15,Houel2012,Kuhlmann2013}. 
Spectral wandering induced by charge noise occurs on a nanosecond timescale, 
as it takes time for charge carriers to accumulate around the QD, and is the dominant dephasing mechanism 
when there is a long delay between excitation pulses. This means that its influence on the coherence properties of 
emitted photons can be heavily suppressed by choosing an appropriately short pulse separation~\cite{Thoma2016,somaschi2016near,Ding2016,Loredo2016,Wang2016}. 
Phonon induced dephasing, on the other hand, has a characteristic timescale of 
picoseconds~\cite{McCutcheon2010,Uns15optica,Wei2014a,Ramsay2010a,McCutcheon2013} and impacts the spectral 
properties of the photons in two principle ways. The first is the emergence of a broad phonon sideband, related to the 
relaxation of the vibrational lattice of the host material during photon emission~\cite{mccutcheon2015optical,Iles-smith2017,Iles-Smith2017_2}. 
The second is a broadening of the zero phonon line due to virtual phonon processes -- here a phonon in the 
material scatters off the QD, driving a virtual transition to an excited electronic state, and 
leading to a random phase change of the emitted photon~\cite{Thoma2016,Muljarov2004,Grange2017}. 

One of the main challenges in the design, engineering and operation of QD single photon sources, lies 
in achieving the maximum reduction of these dephasing processes, which requires an accurate and precise understanding 
of their underlying microscopic origins. In this paper, we present a detailed study of emission from a single 
In(Ga)As QD which is embedded in a highly asymmetric, quasi-planar GaAs low Q-factor microcavity. 
We demonstrate that, while indistinguishable photons with almost ideal interference properties can 
be extracted from this source under resonant excitation conditions, the coherence is strongly affected by 
the pump configuration, the temperature, pump power and the temporal separation between consecutive emission pulses. 
For strictly resonant excitation conditions, our main findings are that a) virtual phonon transitions can strongly affect photon indistinguishability, though have little influence 
on excitonic Rabi oscillations, and that b) photon indistinguishability is independent of excitation pulse area. 
For quasi-resonant excitation, we find noticeably lower indistinguishability values, which further decrease 
with increasing pump power. 
We analyse the experimental data using a model taking into account phonon 
processes both virtual and real in nature, spectral wandering due to charge noise, and timing-jitter from delayed relaxation.

\section{Experimental Setup and Sample description}

\begin{figure}
\begin{center}
\includegraphics[width=0.48\textwidth]{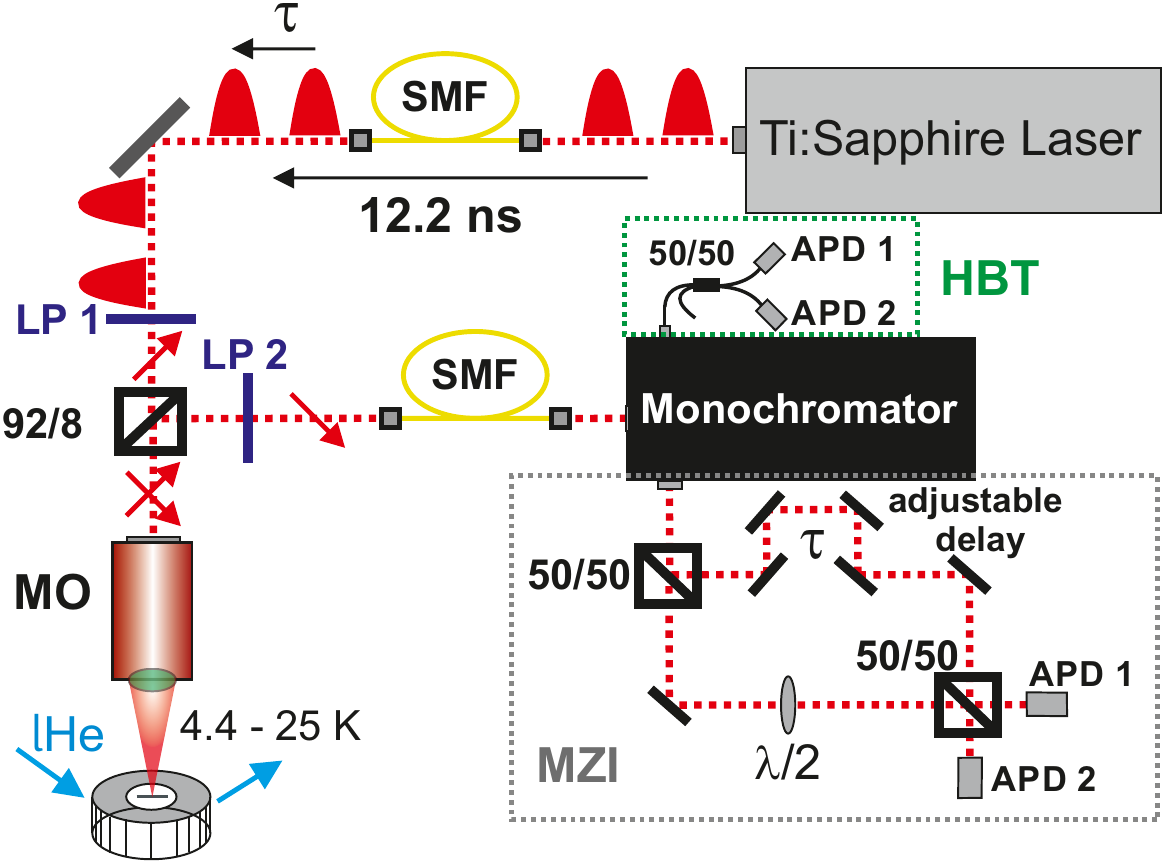}
\caption{Schematic of the experimental setup. The laser pulses separated by 12.2\,ns are each divided into two pulses with 
variable pulse distance $\uptau$. A cross polarization configuration consisting of two linear polarizers (LP 1 \& 2) 
suppresses the scattered laser light by a factor of $\approx$ 10$^7$. After filtering by a single mode fiber and a monochromator, 
the single photons are either directed to the HBT setup (green box) to measure the single photon purity, 
or instead to the unbalanced MZI (grey dotted box) to study photon indistinguishability.}
\label{fig:sketch}
\end{center}
\end{figure}

A sketch of the implemented experimental setup is shown in Fig. \ref{fig:sketch}. 
Short (full width at half maximum of $\delta t\approx 1.2~\mathrm{ps}$) 
coherent laser pulses with a repetition rate 82~MHz are generated 
by a Ti:Sapphire laser, which are each then divided into two separate pulses with an adjustable 
delay $\tau$ by an unbalanced Mach Zehnder Interferometer. 
The spatial pulse shape is fine adjusted by a Single Mode Fiber (SMF) and the polarization is determined by a linear polarizer (LP 1). 
The laser signal is then coupled into the beam path via a 92/8 pellicle beamsplitter and focused on the QD sample by a microscope objective (MO) (NA = 0.42). 
The QD sample is mounted on the cold-finger of a lHe flow cryostat with tuneable temperature. The microscope objective collects the 
emitted QD signal and the second linear polarizer (LP 2), which is orientated perpendicular to the polarization of the laser, 
which suppresses the laser light scattered from the surface of the QD sample. Further filtering processes are accomplished 
by another SMF and by the monochromator with a grating up to 1500\,$\frac{\textnormal{lines}}{\textnormal{mm}}$, 
which is particularly necessary for resonant excitation. As well as suppressing stray excitation light, 
this monochromator spectrally filters the QD signal itself, removing all emission outside a narrow 
$\sim 150~\mu\mathrm{eV}$ window. 

By coupling the filtered QD signal onto a 50/50 beamsplitter, we determine the second order autocorrelation 
function of the source via a standard Hanbury Brown and Twiss (HBT) measurement. To measure emitted photon indistinguishability we 
use an asymmetric Mach--Zehnder interferometer (MZI) to perform a Hong--Ou--Mandel (HOM) interference measurement, 
where the two arms of the interferometer accurately settle the delay between two 
consecutively emitted photons. An additional $\uplambda$/2 wave plate can be inserted to rotate the polarization of the short arm by 
90$^\circ$ to make the two photons distinguishable. In both experiments, the photons 
are detected by two single-photon sensitive silicon based avalanche photo diodes at the exit ports of the second 50/50 beamsplitter. 

The source is composed of a low-density layer of In(Ga)As QDs embedded 
in an asymmetric planar cavity. The resonator consists of five (24) quarter-wavelength
AlGaAs/GaAs mirror pairs in the top (bottom) distributed Bragg
reflector and a 1-$\lambda$-thick central cavity layer. 
High brightness of our source is ensured by oval defects which 
are self-aligned with the QDs and act as natural nano-lenses, enabling an efficiency exceeding 40\,\%~\cite{Maier2014}.

\begin{figure}
\begin{center}
\includegraphics[width=0.48\textwidth]{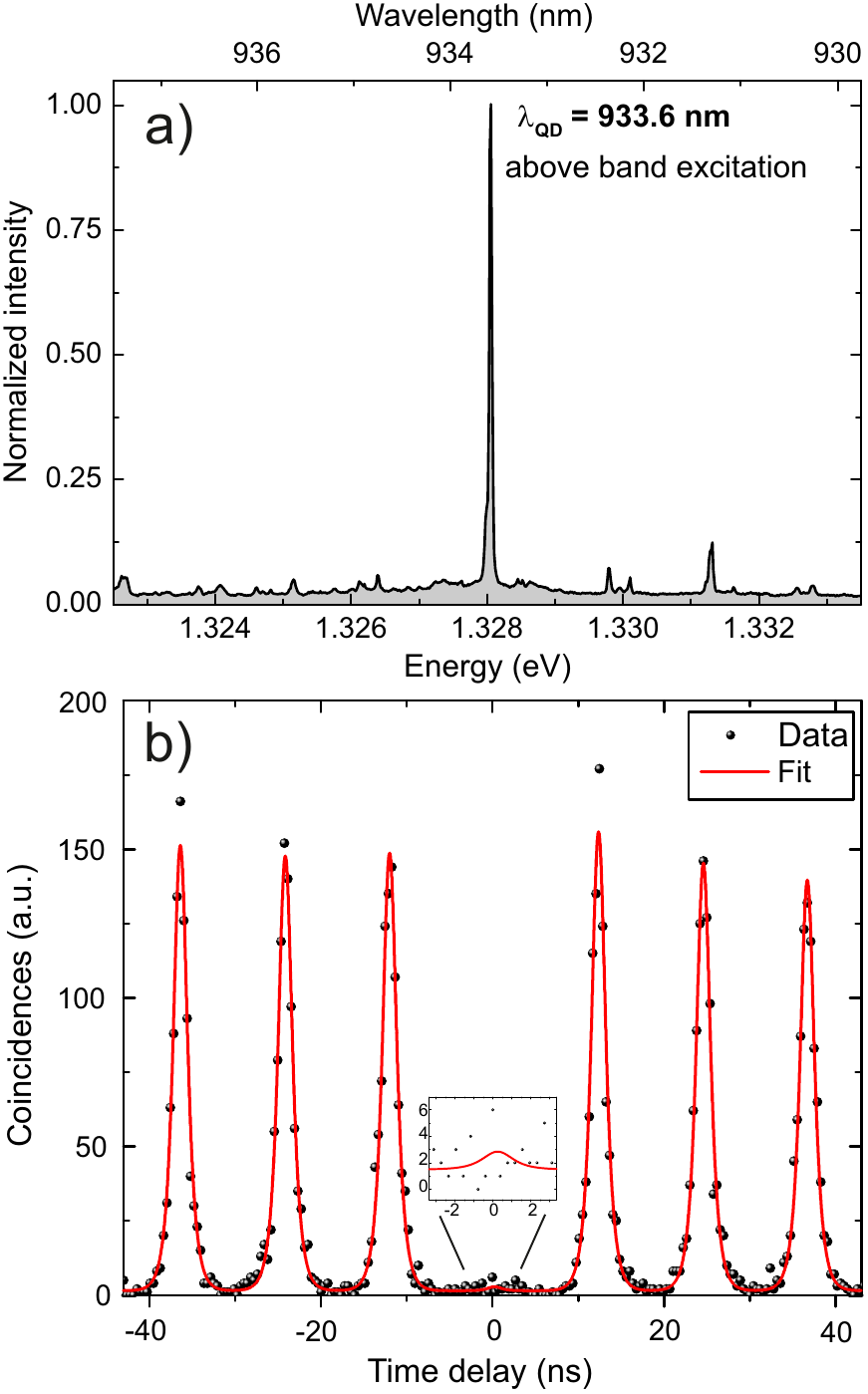}
\caption{(a) Spectrum of the investigated QD using above band excitation. 
(b) Measured second order coincidence histogram under resonant pulsed excitation conditions, 
with pulse separation 12.2\,ns and a $\uppi$-pump power of around 1.6\,$\upmu$W.
The strong lack of coincidence counts around zero delay is an 
unambiguous signature of single photon emission, and we extract $g^{(2)}(0) = 0.006 \pm 0.002$. 
}
\label{fig:spectra}
\end{center}
\end{figure}

Fig. \ref{fig:spectra} (a) shows an above band excitation spectrum of the planar sample with a resolution 
limited QD line at $\lambda_{\mathrm{QD}} = 933.6~\mathrm{nm}$, which we associate with the neutral exciton transition. 
Although one can observe further emission features 
of neighbouring QDs on this scale, the QD line of interest has no distracting emissions in the 
closest energetic vicinity. As we will detail later, in order to obtain a high photon indistinguishability it is indispensable to 
deterministically generate high purity single photons by the use of resonance fluorescence. The measured second 
order autocorrelation function of the QD which can be seen in Fig. \ref{fig:spectra} b) has been carried out under such resonant 
excitation conditions, with the delay between pulses set to their default value of $12.2~\mathrm{ns}$. 
By fitting each pulse of the recorded coincidence histogram with a two sided exponential decay convolved 
with a Gaussian distribution, we find $g^{(2)}(0) = 0.006 \pm 0.002$. This result demonstrates high-purity single photon 
emission of the considered QD. The QD lifetime obtained from these fittings is $T_1$ $\approx$ 730\,ps, which has also been confirmed by time-resolved PL measurements. 

\section{Rabi Oscillations and Phonon Coupling Parameters}

\begin{figure}[tb]
\begin{center}
\includegraphics[width=0.48\textwidth]{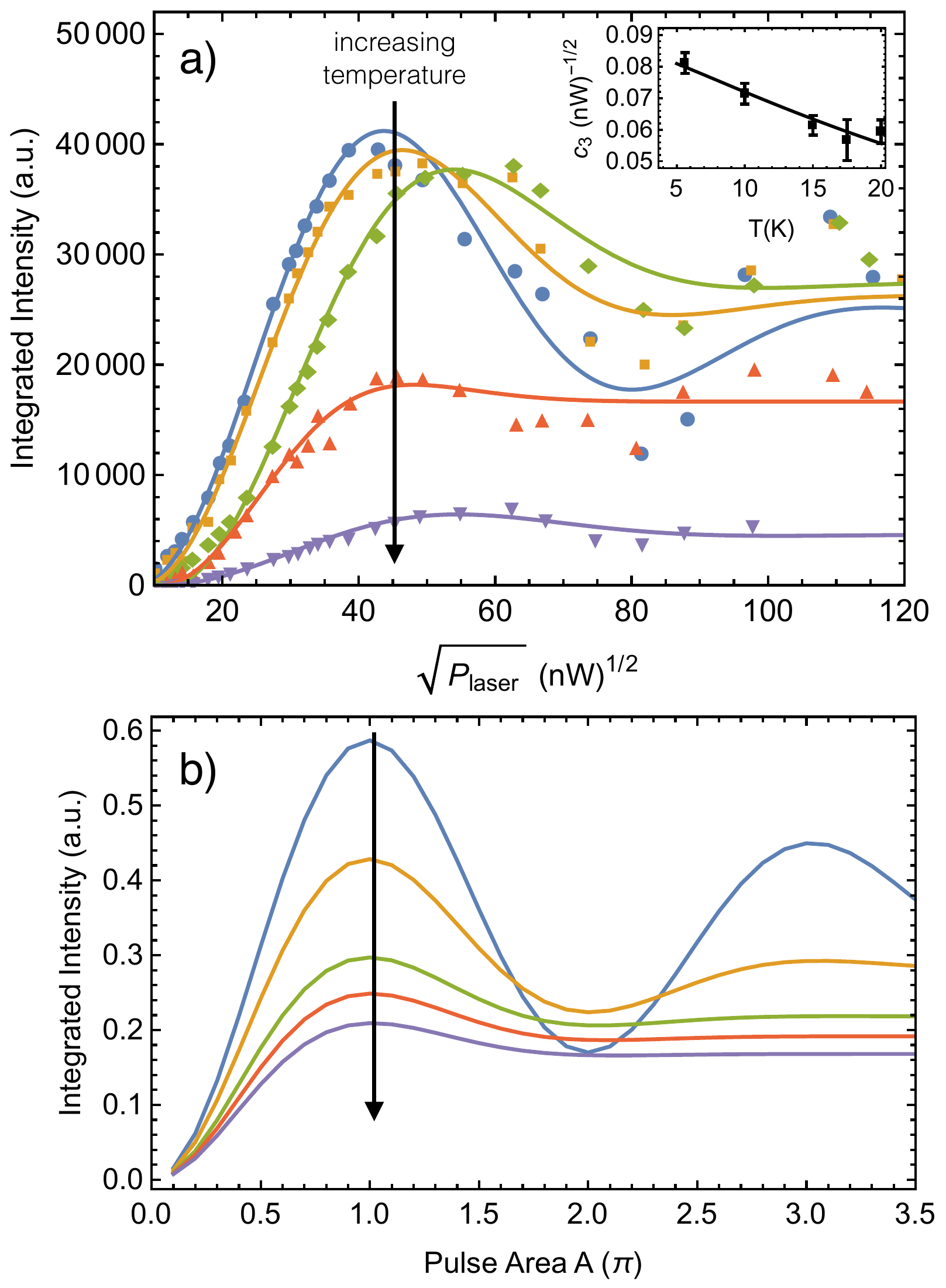}
\caption{a) Rabi-oscillations with increasing pulse power recorded for temperatures of 
$T=5.6,10,15,17.5,20~\mathrm{K}$, ordered as indicated. The fits in 
a) are to a pure-dephasing approximation to the phonon coupling theory. 
The variation of the oscillation period with temperature, shown by the plot markers 
in the inset, allows us to extract 
exciton--phonon coupling parameters, which lead to the solid curve in the inset. 
At higher temperatures, the maximum of the emitted intensity drops, 
and the $\uppi$-pulse shifts towards more intense excitation power. 
b) shows the result of a full phonon coupling model.} 
\label{fig:Rabis}
\end{center}
\end{figure}

Before we study the indistinguishability of the emitted photons, 
it is instructive to first investigate the power and temperature dependent properties of our sample 
through excitonic Rabi oscillations. We excite the QD using  
resonant laser pulses with a fixed temporal width of $\delta t = 1.2$\,ps and a repetition rate of 82\,MHz, and record the integrated intensity as a function 
of the square-root of the laser power, and at various sample temperatures. 
The results are shown in Fig.~\ref{fig:Rabis} for temperatures of 5.6, 10, 15, 17.5 and 20\,K, where damped 
Rabi oscillations are clearly seen~\cite{Ramsay2010,Ramsay2010a,McCutcheon2010,nazir2015modelling,McCutcheon2013}. 
As the laser power is increased, a larger number of 
oscillations between the ground and single-exciton states 
can take place within the pulse duration. 
For increasing  temperature, one can also see a clear 
decrease in overall intensity, as well as an increase in oscillation period.

In order to investigate these features further, we first fit our data to a simplified model of the form 
$ \smash{c_1 [1-\e^{-c_2 A^2}\cos( c_3 A)]}$, where $c_1$, $c_2$ and $c_3$ are constants, 
and $A$ denotes the pulse area, which due to our fixed pulse width satisfies 
$A\propto \smash{P_{\mathrm{laser}}^{1/2}}$~\cite{McCutcheon2013,nazir2015modelling}. 
This form represents a pure-dephasing approximation and makes the simplification of 
temporally flat but finite pulses, such that the integrated intensity can be found from 
known expressions for the exciton excited state population as a function of time~\cite{McCutcheon2010}, 
with the pulse area entering through the dependence on the Rabi frequency $\Omega=A/\delta t$. 
Even within this simplified model, $A$ affects both the period of the oscillations, and also 
the damping through the quadratic dependence, which is a hallmark of excitation-induced 
dephasing~\cite{Ramsay2010,Ramsay2010a,McCutcheon2010}
We note, however, that in general 
coupling to phonons means the dependence of the Rabi oscillations on pulse 
power can be considerably more complex~\cite{McCutcheon2010,Roy-Choudhury:15}. 
Broadly speaking the pure-dephasing approximation is valid when 
$T_1^{-1}\ll\Omega \ll k_B T$ with $\Omega\propto A$ the Rabi frequency. 
Within this approximation the dominant phonon coupling effects are captured 
by a temperature-dependent Rabi frequency renormalisation, here captured in the fitting constant $c_3$, and 
a dephasing rate which goes as the square of the Rabi frequency, hence the exponent in our expression.  

The results of this fitting procedure are shown with the solid curves in Fig.~\ref{fig:Rabis}~(a), 
demonstrating generally good agreement. Despite the simplicity of the present model, we 
can use it to extract important exciton--phonon coupling parameters which we will use in our subsequent analysis. 
Specifically, within the pure-dephasing approximation we use, real phonon transitions cause the Rabi frequency to be renormalised 
by the temperature-dependent Franck--Condon factor $B=\exp[-(1/2)\int_0^{\infty}d\omega\, J(\omega)\omega^{-2}\coth( \omega/(2k_B T))]$~\cite{McCutcheon2010,Iles-smith2017}, 
which in our model is proportional to $c_3$. The renormalisation factor contains the spectral density, 
which we take to be of the form $J(\w)=\alpha \w^3 \exp[-(\w/\w_c)^2]$~\cite{Ramsay2010a,nazir2015modelling}, 
with $\alpha$ an overall exciton--phonon coupling strength and $\w_c$ a phonon cut-off frequency. From the trend of 
$c_3$ with temperature, shown in the inset of Fig.~\ref{fig:Rabis} (a), 
we extract values of $\alpha = 0.13\pm0.01~\mathrm{ps}^{-1}$ and $\w_c=1.8\pm0.1~\mathrm{ps}^{-1}$. 
The values are comparable to those found previously~\cite{Ramsay2010a,nazir2015modelling,Wei2014a,Ulhaq2013}, 
and yield the solid curve shown in the inset.

\begin{figure}[h!]
\begin{center}
\includegraphics[width=0.48\textwidth]{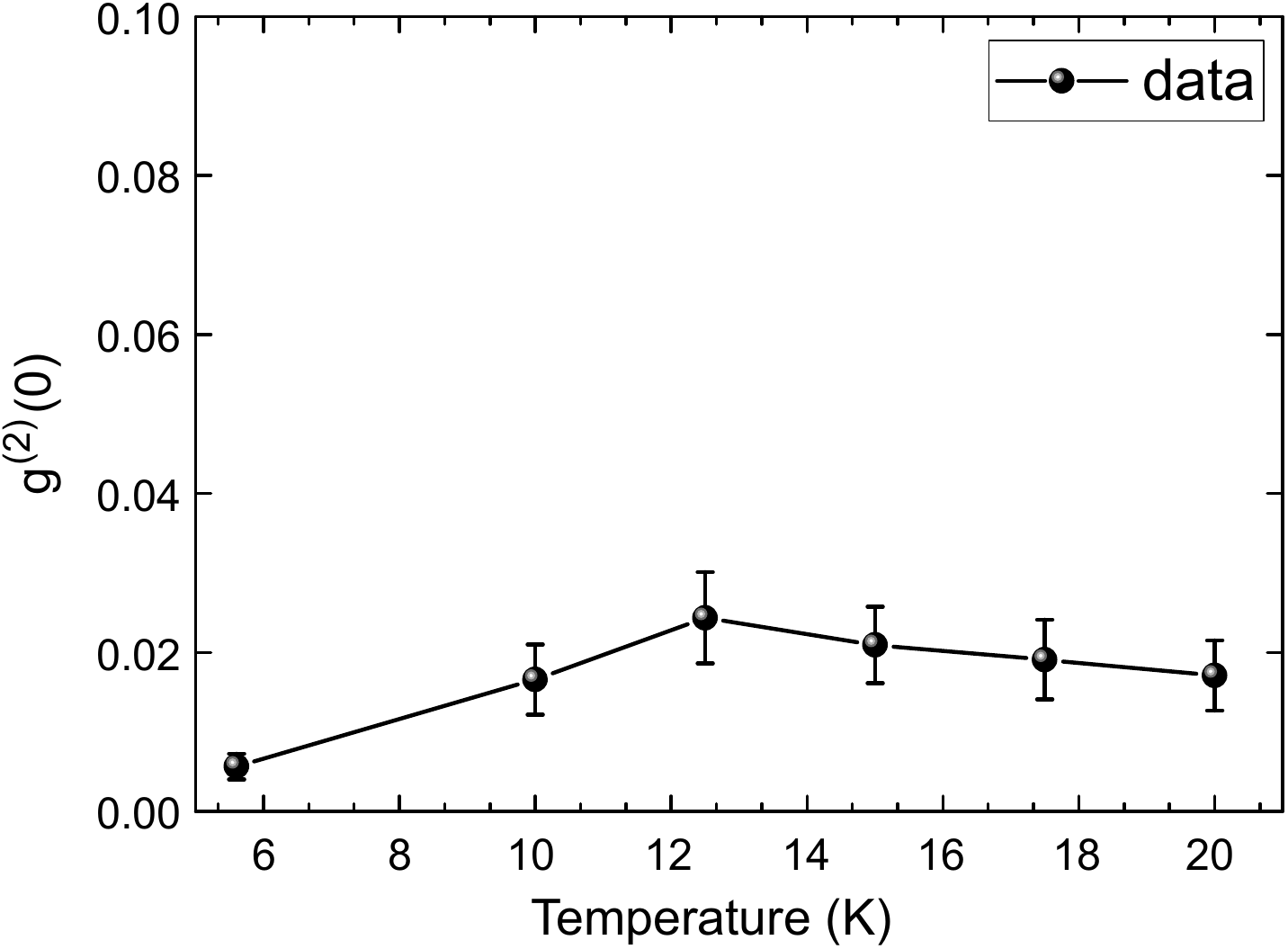}
\caption{Temperature dependent $g^{(2)}(0)$ values, measured at $\uppi$-pulse power each. A slight increase of the multiphoton probability from 0.006 to 0.024 is observable from 5.6 to 12.5\,K, whereas it marginally drops down to $0.017\,\pm\,0.004$ again at 20\,K.}
\label{fig:g2(T)}
\end{center}
\end{figure}

The temperature dependent $g^{(2)}(0)$ values each recorded at the particular $\uppi$-pulse power are presented in Fig.\,\ref{fig:g2(T)}. A slight increase of the multiphoton probability from 0.006\,$\pm$\,0.002 to 0.024\,$\pm$\,0.006 is observable from 5.6 to 12.5\,K, whereas it further marginally drops down to $0.017\,\pm\,0.004$ at 20\,K. The sligthly increased value at 12.5\,K might be induced by an intensified $\uppi$-pulse excitation power, which is nearly doubled in comparison to the 5.6\,K measurement. Except of the distinct lower value of $g^{(2)}(0)$ at 5.6\,K, altogether for our measurements imply that the multiphoton probability is constant for temperatures up to 20\,K. 

\subsection{Virtual Phonon Processes in Rabi Oscillations}

Having partly characterised our system, it is interesting to  
consider whether Rabi oscillations can provide us with any additional characteristics pertinent 
to exciton--phonon coupling in our system. In particular, 
as previously mentioned, phonon interactions in QDs have been shown 
to lead to two primary mechanisms through which photons can lose coherence. 
The first is a coupling induced by displacement of the vibrational lattice due to the changes 
in the charge configuration of the QD~\cite{mccutcheon2015optical,Iles-smith2017,Iles-Smith2017_2}. 
During radiative recombination of an exciton this can lead to the emission or absorption of a lattice excitation in addition to an emitted photon. 
This will produce a shift in the frequency of the emitted photon, leading to the emergence of a broad phonon sideband.

The second process is virtual in nature, in which an incoming phonon drives a virtual transition between the $s$-shell and 
higher lying exciton states in the QD~\cite{Muljarov2004,Reigue2017}. This is a pure-dephasing process where the scattering phonon induces 
a random phase change in the exciton leading to broadening of the zero phonon line (ZPL).  
It is important to note that these two processes have very different temperature dependencies; real transitions occur for all 
temperatures as it is always possible to emit a phonon. Virtual phonon processes, however, require 
non-negligible phonon occupation, and therefore non-zero temperature, in order for a scattering event to occur.  
Thus, as was recently demonstrated by Reigue~\emph{et al.}~\cite{Reigue2017}, we expect real transitions to be the dominant 
dephasing mechanism at low temperatures, and virtual processes to contribute at higher temperatures ($T>10$~K).

In the appendix we derive a master equation describing the laser driven QD exciton, including both phonon 
coupling mechanisms described above. In Fig.~{\ref{fig:Rabis}} (b) we show the results of the full phonon theory, 
for the same temperatures as in (a). We see that the qualitative features are well reproduced, including an overall drop in 
intensity with temperature. This arrises due to an ever greater fraction of phonons being emitted into the 
phonon sideband~\cite{Iles-Smith2017_2}, which in these experiments is removed by filtering. 
Though the overall decrease in maximum intensity with increasing temperature is qualitatively captured by our model, 
we note that there are clear discrepancies. We believe these result from a variation of the sideband with temperature which is not 
quantitatively captured by our model, which assumes a spherically symmetric exciton wavefunction. 
We note that detailed study of the sideband dependence on temperature and QD shape would constitute an interesting and 
relevant study, though is beyond the scope of this work, which instead seeks to explore the behavior of the ZPL under different driving and temperature conditions.

Interestingly, we find that the virtual phonon processes have little impact on 
Rabi oscillations; although these processes are included in Fig.~{\ref{fig:Rabis}} (b), artificially removing them makes 
only imperceptible changes on the scale of the figure. This is because the damping of Rabi oscillations is 
dominated by strong driving induced dephasing, making the real phonon processes orders of magnitude stronger  
than the virtual processes. Hence, although Rabi oscillations can be used to calibrate the real 
phonon-induced processes, they shed no light on the strength of virtual processes and the associated ZPL broadening which 
strongly affects indistinguishability, as we now explore in more detail.

\section{Photon Indistinguishability}

\begin{figure*}[t]
\begin{center}
\includegraphics[width=0.96\textwidth]{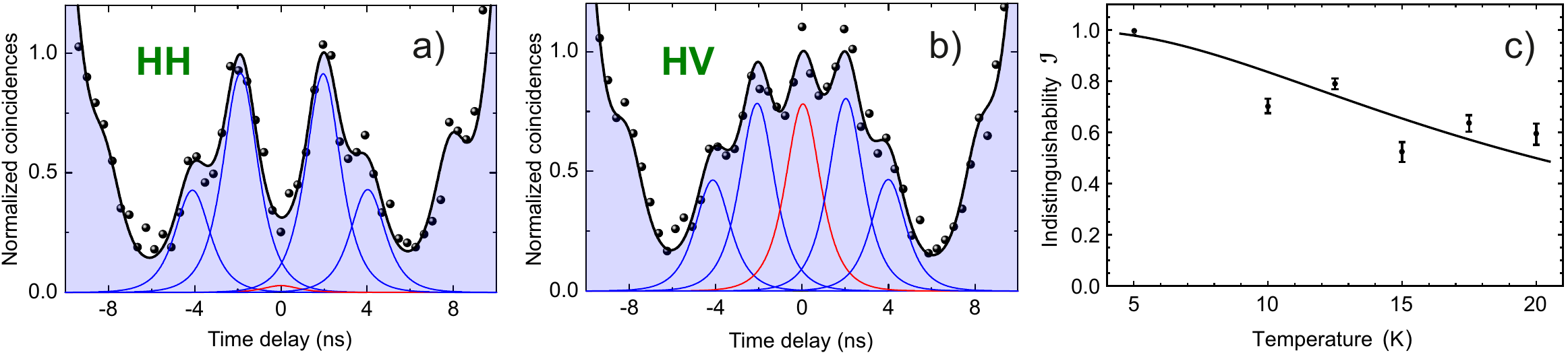}
\caption{Measured two-photon interference histograms for parallel HH (a) and orthogonal photon polarization HV (b). 
The histograms are fitted with a sum of five two-sided exponential functions each convolved with a Gaussian distribution. The measurements were carried out under $\pi$-pulse excitation with an pump power of $P_{Laser} = (1.6 \pm 0.1)\,\mu W$.
These data lead to a corrected indistinguishability of $\mathcal{I} = (99.6\,^{+\,0.4}_{-\,1.4})$\,$\%$. 
(c) Photon indistinguishability as a function of temperature (markers) and theoretical fit (solid line). 
Having spectrally removed the phonon sideband from the QD emission, the temperature dependence here 
arises due to zero-phonon-line broadening attributed to phonon-assisted virtual transitions.}
\label{fig:HH_HV}
\end{center}
\end{figure*}

To measure the indistinguishability of the emitted single photons, we split each excitation pulse into a pair of 
pulses with 2\,ns separation, and then couple these photons into an unbalanced Mach-Zehnder-interferometer. 
Since one arm of the interferometer is precisely adjusted to compensate the delay between the two photons, 
the photons interfere at the second 50/50 beam splitter of the MZI shown in Fig.~\ref{fig:sketch}. 
Figs. \ref{fig:HH_HV} (a) and (b) show the recorded coincidence histograms for 
parallel (HH) and orthogonal polarization (HV) of the photons. We see a near complete suppression 
of the central peak in (a), which is attributed to a high degree of indistinguishability of the emitted single photons. 
To quantitatively assess the degree of indistinguishability, we first calculate  
$\upnu_{\mathrm{raw}}$ = 1\,$\huge-$\,\emph{$(A_{\mathrm{HH}}/A_{\mathrm{HV}})$}, which 
gives the raw visibility, defined in terms of the areas of the central peaks for 
parallel and perpendicular polarisations, and yielding a value of $\upnu_{\mathrm{raw}}$ = (96.3\,$^{+\,0.4}_{-\,1.4}$)\,$\%$. 
Accounting for the measured $g^{(2)}(0)$-value, which slightly deviates from zero ($g^{(2)}(0) = 0.006 \pm 0.002$)
and further taking into account the imperfections of the 50/50 beamsplitter as well as the non-unity contrast of the 
MZI (1\,$\huge-$\,$\upepsilon$ = 0.99), we can correct the raw visibility to determine 
the true single photon indistinguishability~\cite{Santori2002}. 
In doing so we find a corrected indistinguishability 
of $\mathcal{I} = (99.6\,^{+\,0.4}_{-\,1.4})$\,\% \cite{He2013,Wei2014}, taken at $T=5.6~\mathrm{K}$ and 
using strictly resonant $\uppi$ pulses.

\subsection{Dependence on Temperature}

We now explore the photon indistinguishability as described above as a function of sample temperature. 
As seen by the measured data (markers) in Figs. \ref{fig:HH_HV} (c), as temperature is increased, 
photon indistinguishability correspondingly decreases. 
We recall that the monochromator used in our experiment spectrally 
filters the QD emission except for a narrow $\sim 150~\mu\mathrm{eV}$ window around the zero phonon line, 
meaning that the phonon sideband (arising from real phonon transitions and spread over $\sim\mathrm{meV}$) 
is almost entirely removed~\cite{Iles-Smith2017_2}. 
As such, the observed decrease in indistinguishability must arise from ZPL broadening. 

We can therefore conclude that in contrast to Rabi oscillations discussed above, virtual phonon processes do become 
important for indistinguishability measurements as temperature is increased. 
The reason for this difference is related to the different timescales being probed in the two types of measurement. 
For Rabi oscillations induced by temporally short ($\sim 1~\mathrm{ps}$) pulses, only 
strong dephasing processes can have a significant impact. Virtual processes, having typical 
rates of $\sim10^{-4}~\mathrm{ps}^{-1}$ at $T=10~\mathrm{K}$~\cite{Reigue2017} are too weak to affect Rabi oscillations, 
and instead the observed damping is dominated by real phonon processes, 
which take place on a timescsale of $\pi\alpha k_B T (B \Omega)^2\approx 1~\mathrm{ps}^{-1}$ 
at the same temperature~\cite{nazir2015modelling,McCutcheon2013,Wei2014a} 
(for pulse areas of approximately $\pi$ with picosecond pulses such that $\Omega\approx \pi/(1~\mathrm{ps})$).

Indistinguishability measurements, on the other hand, characterise 
emitted photon coherence following QD excitation, 
and are therefore sensitive to processes comparable to the photon emission rate. In the present case, 
$\Gamma = T_1^{-1}\sim 10^{-3}~\mathrm{ps}^{-1}$, allowing relatively weak virtual phonon processes to 
have an impact, as is seen in Figs. \ref{fig:HH_HV} (c). 
This is confirmed by our theoretical model, as shown by 
the solid curve in the plot. Having removed the phonon sideband from the QD emission~\cite{Iles-Smith2017_2}, 
and for very short pulse separations ($\tau_D \leq$ 2 ns, as satisfied here), 
the photon coherence and hence the indistinguishability depends only on the 
emission rate $\Gamma$ and the phonon-induced dephasing rate $\gamma_{\mathrm{pd}}$. 
As shown in the appendix, we use the expression  
\begin{equation}
\mathcal{I} = \frac{\Gamma}{\Gamma + 2\gamma_{\mathrm{pd}}},
\end{equation}
with the temperature dependent rate given by~\cite{Reigue2017,Tighineanu2017}, 
\beq
\gamma_{\mathrm{pd}}= \frac{\alpha^2\mu}{\w_c^4}\int_0^{\infty} \w^{10}\e^{-2(\w/\w_c)^2} n(\w)(n(\w)+1) d\w, 
\label{gammapd}
\eeq
which is a microscopically derived extension of 
the phenomenological expression used in Ref.~\cite{Thoma2016}. 
Three phonon coupling parameters enter these expressions. Using 
the values of $\alpha$ and $\w_c$ already determined from 
the Rabi oscillations in Fig.~{\ref{fig:Rabis}}, we fit the indistinguishability data 
to find the final parameter $\mu=1.1\times 10^{-3}~\mathrm{ps}^2$, 
which is related to the inverse level spacing between s- and p-states of the QD exciton. 
The result of this fitting procedure is shown by the solid curve in Fig. \ref{fig:HH_HV} (c).

\subsection{Dependence on excitation conditions}

We now investigate the influence of the excitation conditions on the photon indistinguishability by varying the 
laser power. Furthermore, and in contrast to previous studies~\cite{Loredo2016}, 
this is done for strictly resonant s-shell excitation, as well as quasi-resonant excitation 
with a QD--laser detuning of $34~\mathrm{meV}$, which is consistent with a p-shell resonance of the QD~\cite{uns15,Grange07}. 
The indistinguishability values were then extracted for 50\,$\%$, 70\,$\%$, and 90\,$\%$ of the QD saturation level under p-shell excitation, 
and for intensities equivalent to $50\,\%$, $75\,\%$ and $100\,\%$ with respect to the maximum of the 
Rabi-oscillation in Fig.~{\ref{fig:Rabis}}~(a) for resonant exciton.

\begin{figure}[t]
\begin{center}
\includegraphics[width=0.48\textwidth]{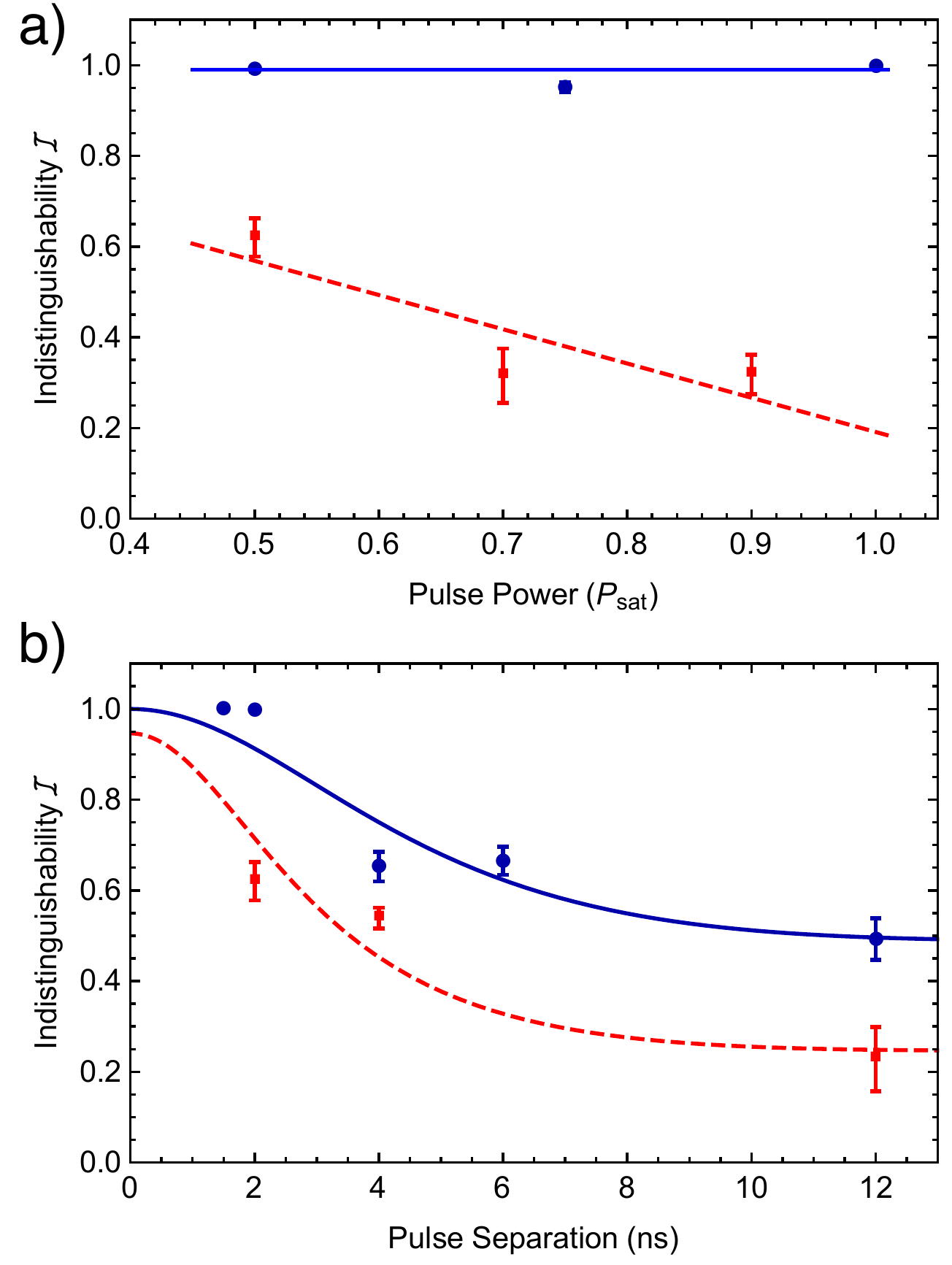}
\caption{a) Indistinguishability as a function of excitation power. For resonant 
s-shell excitation (blue circles), the x-axis corresponds to the pump power as a fraction of a $\pi$-pulse. 
For quasi-resonant p-shell excitation (red squares), we excite $\approx34~\mathrm{meV}$ above the s-shell, 
with the x-axis now corresponding to the fraction with respect to the saturation power. 
b) Indistinguishability as a function of the pulse 
separation. The solid blue curve is a fit to a resonant excitation 
model assuming dephasing only caused by charge fluctuations with a 
finite correlation timescale of $\tau_C\sim 6~\mathrm{ns}$. 
The dashed red curve corresponds 
to a theoretical model also including timing-jitter. Both measurements, a) and b), were carried out at 5.6\,K.}
\label{fig:vis_power}
\end{center}
\end{figure}

The results are presented in Fig. \ref{fig:vis_power} (a). 
Comparing the s- and p-shell data sets, for the latter we can immediately see 
the important role of uncertainties in the relaxation time from p- to s-shell (timing-jitter), 
and also the increased probability of exciting charge carriers in the vicinity of the QD (giving rise to dephasing), 
both of which suppress indistinguishability~\cite{Kaer2013,uns15}. 
Additionally, see that compared to the resonant pumping conditions, 
which give a near-unity degree of indistinguishability 
for all chosen pump powers, the indistinguishability for p-shell excitation significantly drops down from 
$\mathcal{I}$ = (62 $\pm$ 4)\,$\%$ for $P=0.5P_{\mathrm{sat}}$, 
to $\mathcal{I}$ = (32 $\pm$ 4)\,$\%$ for $P\approx0.9P_{\mathrm{sat}}$. 

Timing-jitter is parameterised 
by a single decay constant characterising the p- to s- shell transition, which being 
phonon mediated, is expected to be independent of optical excitation power~\cite{Grange07}. 
The decrease of indistinguishability with increasing power, as has been observed elsewhere~\cite{Loredo2016}, 
gives us an indication as to the relative strength of timing-jitter and dephasing caused by charge fluctuations.  
Although a complete set of power dependent measurements (including 
very low powers) would be necessary to fully determine the relative strengths of 
these two mechanisms, we can nevertheless assume that the dephasing contribution 
leads to a linear decrease in indistinguishability with power, as was found in Ref.~\cite{Loredo2016}. 
The result of this fitting procedure is shown by the dashed red curve, and by extrapolating 
to zero power we find $\mathcal{I}\to 95\,\%$. Within these assumptions, at zero power 
the value of $\mathcal{I}$ is now dominated by timing-jitter, and given by 
$\mathcal{I}=\Gamma_{p\to s}/(\Gamma_{p\to s}+\Gamma)$, where 
$\Gamma_{p\to s}^{-1}=53~\mathrm{ps}$ is the p- to s-shell relaxation time.

Turning now to strictly resonant excitation, interestingly there is little change in the measured 
indistinguishability as a function of power. This suggests that the power-dependent dephasing processes 
present for p-shell excitation are now suppressed, and 
also that the power dependent dephasing 
observed in the Rabi oscillations in Fig.~\ref{fig:Rabis} does not affect photon indistinguishability. 
The apparent absence of any power dependent dephasing can be explained 
by noting that the saturation power for s-shell excitation is significantly smaller than 
for p-shell excitation, and one may therefore expect the creation of charge carriers to be correspondingly suppressed. 
Additionally, the insensitivity of indistinguishability measurements to the phonon-induced damping seen in Fig.~\ref{fig:Rabis} 
can be explained by noting that phonons lead to thermalisation in the dressed state basis, mediating population transfer between 
the system eigenstates, and giving rise to a reduced population in the single exciton state. 
Measurements of indistinguishability, however, are normalised in such a 
way as to naturally post-select events where the 
QD has been successfully excited, removing any dependence on the initial QD population. 
More generally, as we show in the 
appendix, for very short excitation pulses satisfying $\delta t/T_1 \ll1$, indistinguishability 
is independent of the QD state immediately following excitation.

\subsection{Dependence on pulse separation}

While phonons affect the coherence of QD excitons on a timescale which is short compared to the exciton lifetime, 
the effects of fluctuating charges and spin noise occur on a nanosecond timescale~\cite{Thoma2016}. 
Thus, in order to quantify the impact of these slower channels on the photon interference, it is instructive to study the 
indistinguishability as a function of the pulse separation between the arriving excitation pulses. 
To do so we vary the pulse separation between 2 ns and 12 ns, which maps out 
the characteristic correlation time of dephasing processes which act on the source. This is repeated 
for resonant ($\pi$-pulse conditions) as well as quasi-resonant excitation conditions 
(50 $\%$ of the saturation level).

The results are plotted in Fig. \ref{fig:vis_power} (b), where in both cases we observe a  
drop in the indistinguishability as the pulse separation is increased. 
In order to gain a quantitative understanding, we make use of the expression developed by Thoma~\emph{et al.}~\cite{Thoma2016}, 
for which charge and spin noise are assumed to lead to a stochastic fluctuation of the excited state energy level, and taken to 
have Gaussian statistics. By taking an ensemble average at a given pulse separation $\tau_D$ 
over the noise correlation function, a Gaussian form for the associated dephasing rate can be obtained as 
\begin{equation}
\tilde\gamma(\tau_D) = \tilde\gamma_0\left(1-e^{-(\tau_D/\tau_C)^2}\right),
\label{tildegamma}
\end{equation} 
where $\tau_C$ quantifies the finite correlation time of the environment, 
and $\tilde{\gamma}_0$ parameterises the overall strength of fluctuations. 
In general, this rate should be added to the temperature dependent contribution in 
Eq.~({\ref{gammapd}}), though in the following this latter contribution is ignored due 
to the low temperature of $T=5.6~\mathrm{K}$ used. We therefore 
use simply $\mathcal{I}=\Gamma/(\Gamma+2\tilde{\gamma}(\tau_D))$ to fit the 
s-shell excitation data in Fig. \ref{fig:vis_power} (b), with the results shown by the solid blue curve. 
The fitting parameters we find are $\tilde{\gamma}_0=0.37~\mu\mathrm{eV}$ and 
$\tau_C=6.48~\mathrm{ns}$, which give $\mathcal{I}=0.49$ as $\tau_D\to\infty$.  

In the p-shell excitation case the situation is somewhat more complicated. 
Even in the absence of any phonon-induced dephasing, we expect both timing-jitter 
and a power dependent dephasing rate to reduce 
phonon indistinguishability~\cite{Kaer2013,uns15}. 
We can, nevertheless, use the p- to s-shell relaxation rate $\Gamma_{p\to s}$ found from 
Fig. \ref{fig:vis_power} (a), and again assume that all the dephasing mechanisms causing a loss 
in indistinguishability are described by Eq.~({\ref{tildegamma}}). 
Including both timing-jitter and dephasing, we fit the data in Fig. \ref{fig:vis_power} (b) to
\beq
\mathcal{I}=\left(\frac{\Gamma_{p\to s}}{\Gamma_{p\to s}+\Gamma}\right)\left(\frac{\Gamma}{\Gamma+2\tilde{\gamma}(\tau_D)}\right),
\label{IWithTJ}
\eeq
with the result shown by the dashed red curve, and we find fitting parameters 
$\tilde{\gamma}_0=1.0~\mu\mathrm{eV}$ and $\tau_C=5.8~\mathrm{ns}$. 
As is expected from Fig. \ref{fig:vis_power} (a), s- and p-shell excitation 
conditions give rise to different dephasing rates $\tilde{\gamma}$. 
More intriguingly, by comparison of Figs. \ref{fig:vis_power} (a) and (b), 
it appears this dephasing process behaves similarly as a function of time (pulse separation), 
but quite differently as a function of power. 
Whether the dephasing taking place in each case is of the same origin remains to be seen, 
and it would be interesting to investigate the entire 
pump power--pulse separation parameter space in future studies.

\section{Conclusion}

We have studied both Rabi oscillations and photon indistinguishability for a QD in a 
quasi-planar low-Q microcavity. For resonant excitation of a single exciton, 
we found that increasing temperatures strongly affected indistinguishability, 
though had little influence on excitonic Rabi oscillations. With increasing pump 
power, we found that indistinguishability was unaffected in the resonant case, 
though was strongly reduced for quasi-resonant excitation conditions even 
for short pulse separations, suggesting a fast dephasing process acting on a 
timescale shorter than $2~\mathrm{ns}$. With increasing pulse separation, 
indistinguishability dropped in both the resonant and quasi-resonant cases, 
from which we deduce a characteristic noise correlation time of 
approximately $6~\mathrm{ns}$.

\begin{acknowledgements}
We acknowledge support by the State of Bavaria and the German Ministry of Education and Research (BMBF) within the project Q.com. 
J.I.-S. and J.M. acknowledge support from the Danish Research Council (DFF-4181-00416) and 
Villum Fonden (NATEC Centre). This project has received funding from the 
European Union's Horizon 2020 research and innovation programme under the 
Marie Sk{\l}odowska-Curie grant agreement No. 703193.
\end{acknowledgements}

\begin{appendix}
\section{Exciton--Phonon Coupling Model}
\label{appendix}

We model the QD as a two-level system with ground-state $\ket{0}$ and first exciton state $\ket{X}$, 
corresponding to the $s$-shell transition of the QD with energy $\omega_X$. 
The excitation is consider to be a semi-classical laser pulse with carrier frequency $\omega_l$, and a Gaussian envelope 
function $\Omega(t) = (A/2\delta\tau\sqrt{\pi}) \exp(-(t-t_0)^2/(2\delta\tau)^2)$, where $A$ is the pulse area, 
and $\delta\tau$ is the temporal pulse width, such that $\delta\tau=\delta t /(4\sqrt{\ln 2})$ with $\delta t$ the full width half maximum. 
By moving to a frame rotating with respect to the carrier frequency and making the rotating wave approximation, 
we obtain a time-dependent system Hamiltonian $H_S(t) = \frac{\delta}{2}\sigma_z + \frac{\Omega(t)}{2}\sigma_x$, 
where we have defined the detuning $\delta = \omega_X - \omega_L$. The Pauli operators take their standard form 
$\sigma_x = \ketbra{0}{X} +\ketbra{X}{0}$, $\sigma_y = i(\ketbra{0}{X} - \ketbra{X}{0})$, and $\sigma_z = \ketbra{X}{X} - \ketbra{0}{0}$.

To describe the influence of phonon dephasing mechanisms, 
we model the phonon environment as a collection of harmonic oscillators with free Hamiltonian 
$H_B = \sum_{\vk}\nu_{\vk} b^\dagger_{\vk}b_{\vk}$, where $b_{\vk}$ is the annihilation operator for the $\vk^{th}$-mode of the phonon environment with frequency $\nu_{\vk}$.  
The electron-phonon interaction is then governed by the Hamiltonian  $H_I = \ket{X}\bra{X} (V_L + V_Q)$. 
The first term, $V_L = \sum_{\vk} g_{\vk} (b_{\vk}^\dagger + b_{\vk})$, leads to real phonon dephasing mechanisms 
as well as the emergence of the phonon sideband~\cite{Iles-smith2017,Reigue2017,Iles-Smith2017_2}. 
The linear electron-phonon coupling strength is given by the matrix elements $g_{\vk} = \sum_{a = e,h} M_{a,\vk}^{11}$ 
for electrons (e) and holes (h), where for deformation potential coupling we have:
$$
M_{a,\vk}^{ij} = \sqrt{\frac{\nu_{\vk}}{2\varrho c_s^2\mathcal{V}}}D_a\int d^3r\psi_{i a}^\ast(\vecr)\psi_{j a}(\vecr)e^{i\vk\cdot\vecr},
$$
which is the matrix element corresponding to phonon induced transition between the $i^{th}$- and $j^{th}$ electronic states. 
Here $\varrho$ is the mass density, $c_s$ is the speed of sound in the material, and $\mathcal{V}$ is the phonon normalisation volume. 
This matrix element depends on the wavefunction $\psi_{i,e/h}(\vecr)$ of the confined electron/hole 
and the corresponding deformation potential $D_a$. 

As introduced by Muljarov and Zimmerman~\cite{Muljarov2004}, virtual phonon-assisted processes can be described by a quadratic interaction term 
$V_Q = \sum_{\vk,\vk^\prime} f_{\vk,\vk^\prime}(b_{\vk}+b^\dagger_{\vk})(b_{\vk^\prime}+b^\dagger_{\vk^\prime})$, 
where higher lying states are eliminated perturbatively, allowing us to treat the QD as a two level system. 
The effective coupling for the virtual processes is given by 
$f_{\vk,\vk^\prime} = \sum_{a=e,h}\sum_{j>1}M^{1j}_{a,\vk}M^{j1}_{a,\vk^\prime}[\omega^{a}_m - \omega^a_1]^{-1}$, 
where $\omega^{e/h}_m$ is the energy of the $m^{th}$ electron/hole energy level.

\section{Rabi oscillation master equation}
To describe the effect of phonon interactions on the exciton dynamics, we shall use a 
polaron master equation approach~\cite{McCutcheon2010,manson2016polaron,Iles-smith2017,Reigue2017}. 
Here we apply the unitary transformation $\mathcal{U}_V = \exp\left(-\sigma^\dagger\sigma \otimes S\right)$, 
with $S = \sum_{\vk} \nu_{\vk}^{-1}g_{\vk}(b_{\vk}^\dagger - b_{\vk})$ to the electron-phonon coupling Hamiltonian defined above.
This leads to a displaced representation of the phonon environment, providing an optimised basis for a perturbative description of the QD dynamics. 
Importantly, this transformation naturally captures the non-Markovian relaxation behaviour of the phonon environment 
during exciton recombination~\cite{Iles-smith2017,Reigue2017,Iles-Smith2017_2}.

Applying the polaron transformation to the Hamiltonian yields $H_V = \mathcal{U}_V^\dagger H\mathcal{U}_V=  H_S  + H_I + H_B$, where
\begin{equation}\begin{split}
&H_S = \frac{\tilde\delta}{2}\sigma_z + \frac{\Omega_r(t)}{2}\sigma_x,~~H_B=  \sum\limits_{\vk}\nu_{\vk}b^\dagger_{\vk}b_{\vk},\\
& H_{I} = \frac{\Omega(t)}{2}(\sigma_x B_x + \sigma_y B_y) + \ketbra{X} {X}V_Q. 
\end{split}
\end{equation} 
Here $\tilde\delta = \delta + \sum_{\vk}\nu_{\vk}^{-1}g_{\vk}^2$ is the polaron shifted detuning, and the driving term 
$\Omega_r(t) = B\Omega(t)$ has been renormalised by the thermal expectation of the lattice displacement operator 
$B = \tr\left(\exp(\pm S)\rho_{B}\right)$ and $\rho_B = \exp\left(-\beta H_B\right)/\tr\left[\exp\left(-\beta H_B\right)\right]$, 
with the thermodynamic temperature defined as $\beta = 1/k_B T$.
In this transformed representation, the system now couples to the phonon environment through the displacement operators $B_x = (B_+ + B_- - 2B^2)/2$, 
and $B_y = i(B_+-B_-)/2$ with $B_{\pm} = \exp(\pm S)$.

This Hamiltonian allows us to derive a second-order master equation in the polaron frame which captures multi-phonon processes~\cite{McCutcheon2010}. 
 Using the Born-Markov approximation~\cite{breuer2002theory}, we may write the Schr\"odinger picture master equation as: 
\begin{widetext}
\begin{equation}
\begin{split}
&\frac{\partial\rho(t)}{\partial t} = -i\left[\frac{\delta}{2}\sigma_z + \frac{\Omega_r(t)}{2}\sigma_x,\rho(t)\right]
+\int\limits_0^\infty d\tau\left( \left[\sigma^\dagger\sigma,\sigma^\dagger\sigma(t-\tau,t)\rho(t)\right] \Lambda_z(\tau)+ \operatorname{h.c.}\right)\\
&+\frac{\Omega(t)}{4}\int\limits_0^\infty d\tau \Omega(t-\tau)\left(\left[\sigma_x,\sigma_x(t-\tau,t)\rho(t)\right]\Lambda_x(\tau)+\left[\sigma_y,\sigma_y(t-\tau,t)\rho(t)\right]\Lambda_y(\tau) +\operatorname{h.c.}\right),
\end{split}
\end{equation}
\end{widetext}
Here we have defined $\sigma_i(s,t) = U_0(t)U_0^\dagger(s)\sigma_i U_0(s)U_0^\dagger(t)$, where $U_0(t) = \exp\left( -i\int_0^t H_S(s)ds\right)$ is the interaction picture transformation.
The polaronic correlation functions take the standard form~\cite{McCutcheon2010}:
\begin{equation}\begin{split}
&\Lambda_x(\tau) = \frac{B^2}{2}\left(e^{\varphi(\tau)} + e^{-\varphi(\tau)}-2\right),\\
&\Lambda_y(\tau) = \frac{B^2}{2}\left(e^{\varphi(\tau)} - e^{-\varphi(\tau)}\right),
\end{split}\end{equation}
where we have defined the phonon propagator as $\varphi(\tau) = \int_0^\infty d\nu  \nu^{-2}J(\nu)\left(\cos\nu\tau\coth\left(\beta\nu/2\right) - i\sin\nu_k\tau\right)$. 
Here the spectral density takes the standard form $J(\nu) = \alpha \nu^3\exp(-\nu^2/\nu_c^2)$ where 
$\alpha$ is the linear electron-phonon coupling strength, set by the deformation potential, and $\nu_c$ is the cut-off frequency, 
set by the size of the exciton wavefunction~\cite{nazir2015modelling}.

In addition to the polaronic dephasing terms, we also have a correlation function associated with the virtual transitions: 
$\Lambda_z(\tau) = \sum_{\vk,\vk^\prime} \sum_{\bm{q	},\bm{q}^\prime}f_{\vk,\vk^\prime}f_{\bm{q	},\bm{q}^\prime}
\langle (b_{\vk}(\tau)+b_{\vk}^\dagger(\tau))(b_{\vk^\prime}(\tau)+b_{\vk^\prime}^\dagger(\tau))(b_{\bm{q}}
+b_{\bm{q}}^\dagger)(b_{\bm{q}^\prime}+b_{\bm{q}^\prime}^\dagger)\rangle$. 
For a QD with a spherically symmetric wavefunction the primed and un-primed modes may be factorised -- 
this is equivalent to assuming that the incoming and going scattering modes are different~\cite{Reigue2017}. 
Assuming that the dominant virtual transition is between the $s-$ and $p-$shells, we find that 
\begin{equation*}\begin{split}
&\Lambda_z(\tau) = \left[\int\limits_0^\infty d\omega \mathcal{J}(\omega)\left(n(\omega) e^{i\omega\tau} + (n(\omega) + 1)e^{-i\omega\tau}\right)\right]^2,
 \end{split}\end{equation*}
where we have defined the thermal occupation as $n(\omega) = [\exp(\beta\omega)-1]^{-1}$.
We have also introduced 
$\mathcal{J}(\omega) = \sqrt\alpha_Q \omega^5 \exp({-\omega^2/\omega_c^2})$, 
where  $\alpha_Q = \alpha^2 \omega_c^{-4}\mu$ is the virtual phonon coupling strength. 
Here $\mu = \pi[D_e-D_h]^{-4}(D_e^2\Delta_e^{-1}+D_h^2\Delta_h^{-1})$ and $\Delta_{e/h}$ is the splitting between the s- and p-shells for the electron/hole. 
A detailed derivation of this expression can be found in the supplement of Ref.~\cite{Reigue2017}.

In the Markov approximation, the phonon correlation function decays on a timescale much faster than the system dynamics.
This allows us to make an adiabatic approximation when transforming to the interaction picture~\cite{McCutcheon2010}, 
such that $\sigma_i(t-\tau)\approx \exp(-iH_s(t)\tau)\sigma_i \exp(iH_s(t)\tau)$. 
On resonance with polaron shifted transition energy, $\tilde{\delta}=0$, 
the interaction picture system operators then take the form
\begin{equation*}\begin{split}
\sigma_x(t-\tau,t) = &\sigma_x,\\
\sigma_y(t-\tau,t) = &\sin(\Omega_r(t)\tau)\sigma_z + \cos(\Omega_r(t)\tau)\sigma_y,\\
\sigma^\dagger\sigma(t-\tau,t) =&\frac{1}{2}\left(1 - \cos(\Omega_r(t)\tau)\right)\mathbb{1}\\
+&\cos(\Omega_r(t)\tau) \sigma_z +\frac{1}{2} \sin(\Omega_r(t)\tau)\sigma_y.
\end{split}\end{equation*}
Substituting these expressions into our master equation, we have:
$\dot\rho(t)= -\frac{i\Omega_r(t)}{2}\left[\sigma_x,\rho(t)\right]  + \mathcal{K}_{L}[\rho(t)]+ \mathcal{K}_{Q}[\rho(t)]$ 
where we have defined the super-operator associated to real phonon transitions as 
\begin{equation*}\begin{split}
&\mathcal{K}_L[\rho(t)] = -\left(\frac{\Omega(t)}{2}\right)^2\left(\left[\sigma_x,\sigma_x\rho(t)\right]\Gamma_1(t)\right.\\
&\left. +\left[\sigma_y,\sigma_y\rho(t)\right]\Gamma_2(t)
 +\left[\sigma_y,\sigma_z\rho(t)\right]\Gamma_3(t) +\operatorname{h.c.}\right),
\end{split}\end{equation*}
and those associated with virtual transitions as 
\begin{equation*}\begin{split}
&\mathcal{K}_Q[\rho(t)] =\\
&\left[\sigma^\dagger\sigma,(\chi_1(t)+\chi_{2}(t)\sigma_y + \chi_3(t)\sigma^\dagger\sigma)\rho(t)\right] 
+\operatorname{h.c.}.
\end{split}\end{equation*}
By defining the Fourier transformed correlation function $\gamma_i(\omega) = \int_0^\infty d\tau\Lambda_i (\tau) \exp(i\omega\tau) )$, 
we may write the rate functions for real transitions as $\Gamma_1 = \gamma_x(0)$, $\Gamma_2(t) = [\gamma_y(\Omega_r(t)) + \gamma_y(-\Omega_r(t))]/2$, and 
$\Gamma_3(t) = [\gamma_y(\Omega_r(t)) - \gamma_y(-\Omega_r(t))]/2i$. 
The rates associated with the virtual transitions are then $\chi_{1}(t) = \gamma_z(0)/2 - [\gamma_z(\Omega_r(t))+\gamma_z(-\Omega_r(t))]/4$, 
$\chi_2(t) = [\gamma_z(\Omega_r(t)) - \gamma_z(-\Omega_r(t))]/4i$, and $\chi_3(t) = [\gamma_z(\Omega_r(t)) + \gamma_z(-\Omega_r(t))]/2$.

As discussed in the main text, the Rabi oscillation measurements probe picosecond timescales. 
Virtual phonon processes, on the other hand, occur on a nano-second timescale, therefore the impact of the pulse on the virtual phonon dissipator is negligible. 
This allows us to replace the time dependent dissipator $\mathcal{K}_Q[\rho(t)]$ with its time independent counterpart, that is 
$$
\mathcal{K}_Q [\rho(t)]\approx \frac{\gamma_{\mathrm{pd}}}{2}\mathcal{L}_{\sigma^\dagger\sigma}[\rho(t)],
$$
where $\mathcal{L}_O[\rho] = 2O\rho O^\dagger - \left\{O^\dagger O,\rho\right\}$, and 
$\gamma_{\mathrm{pd}} = \int_0^\infty d\omega \mathcal{J}(\omega)^2n(\omega)(n(\omega)+1)$, 
is the pure dephasing rate due to the virtual phonon scattering~\cite{Reigue2017,Tighineanu2017}.
Using this expression, the final master equation becomes 
\begin{equation}\begin{split}
\frac{\partial\rho(t)}{\partial t} \approx &-\frac{i\Omega_r(t)}{2}\left[\sigma_x,\rho(t)\right]  + \mathcal{K}_{L}[\rho(t)] \\
&+ \frac{\gamma_{pd}}{2}\mathcal{L}_{\sigma^\dagger\sigma}[\rho(t)].
\label{RabiME}
\end{split}
\end{equation}

By numerically solving Eq.~({\ref{RabiME}}) we are able to explore how the final exciton population 
behaves as a function of pulse area and temperature. The final ingredient necessary to obtain 
a value proportional to the measured integrated intensity is the fraction of emission events 
into the zero phonon line. As shown in Ref.~\cite{Iles-Smith2017_2} this is simply the 
square of the thermal expectation of the lattice displacement operator, $B$. As such, 
for Fig.~\ref{fig:Rabis} (b) we solve Eq.~({\ref{RabiME}}) for the excited state population 
and multiply by $B^2$. 

\section{Indistinguishability calculations}

Hong-Ou-Mandel measurements probe the coherence properties of photons on time-scales relevant to the emitter lifetime, in this case $T_1\sim10^{3}$~ps. 
This is long after the pulse has finished, such that $\mathcal{K}_L[\rho(t)]\rightarrow0$, meaning that the only phonon dephasing processes remaining 
in the master equation are virtual in nature; the master equation reduces to
\begin{equation}\begin{split}
\frac{\partial\rho(t\gg\delta\tau)}{\partial t} =& \frac{\gamma}{2}\mathcal{L}_{\sigma^\dagger\sigma}[\rho(t)]+ \frac{\Gamma}{2}\mathcal{L}_{\sigma}[\rho(t)]. 
\label{eq:PDmeq}
\end{split}\end{equation}
Here $\gamma = \gamma_{\mathrm{pd}} + \tilde{\gamma}(\tau_D)$, where $ \tilde\gamma(\tau_D)$ 
is introduced to capture dephasing associated with charge noise~\cite{Thoma2016}, and is given in Eqn.~({\ref{tildegamma}}). 

If the phonon sideband has been removed from the QD spectrum~\cite{Iles-Smith2017_2}, as is the case in our experiments, 
the indistinguishability can be found directly from the first order 
correlation function $g^{(1)}(t,t+\tau)=\langle \sigma^{\dagger}(t+\tau)\sigma(t)\rangle$ via the expression~\cite{uns15,Kaer2013}
\begin{equation}
\mathcal{I} = \frac{\int_0^\infty dt \int_0^\infty d\tau \vert g^{(1)}(t,t+\tau)\vert^2 }{\int_0^\infty dt\int_0^\infty d\tau g^{(1)}(t,t)g^{(1)}(t+\tau,t+\tau)}. 
\label{Iwithg}
\end{equation}
By using the quantum regression theorem~\cite{carmichael1999statistical}, we can express the correlation 
function in terms of the QD density operator $g^{(1)}(t,t+\tau) =\rho_{XX}(t) \exp(-(\Gamma + 2\gamma)\tau/2)$, 
where $\rho_{XX}(t) = \rho_{XX}(0)\exp(-\Gamma t/2)$ is the time evolution of exciton population, with $\rho_{XX}(0)$ the excitonic population 
immediately following the excitation pulse. 
It is clear from the form of this correlation function that it is independent 
of the exciton coherence established by the pulse, and depends only on the exciton population.
Thus, in situations where $T_1\gg\delta\tau$, the driving dependent dephasing observed in the 
Rabi oscillations does not impact the indistinguishability. Furthermore, all occurrences of 
$\rho_{XX}(0)$ in Eq.~({\ref{Iwithg}}) will cancel, and we can say more generally that the indistinguishability 
is independent of the initial exciton state. Eqs.~({\ref{eq:PDmeq}}) and ({\ref{Iwithg}}) lead to the indistinguishability 
expressions used in the main text. 

%\subsection{Timing jitter}

When considering $p-$shell excitation of the QD we must account for the timing-jitter introduced by finite relaxation rate 
to the single exciton $s$-shell from which photon emission occurs. 
To do so we introduce a third `pump' state $\ket{P}$~\cite{uns15} with energy $\Delta$, 
which decays to the $s$-shell state $\ket{X}$ with rate $\Gamma_{p\rightarrow s}$. 
The master equation describing this process, in addition to charge noise and virtual phonon scattering, is given by 
\begin{equation}\begin{split}
\frac{\partial\rho(t)}{\partial t} =&  - i \left[\Delta \ket{P}\bra{P},\rho(t)\right]+\frac{\gamma}{2}\mathcal{L}_{\sigma^\dagger\sigma}[\rho(t)]\\
&+ \frac{\Gamma}{2}\mathcal{L}_{\sigma}[\rho(t)] + \frac{\Gamma_{p\rightarrow s}}{2}\mathcal{L}_{\ket{X}\bra{P}}[\rho(t)],
\label{eq:PDmeq2}
\end{split}\end{equation}
Following the same procedure as before, though initialising the QD in $\ket{P}$, we obtain the modified 
expression for the indistinguishability~\cite{uns15,Kaer2013}:
$$
\mathcal{I} =\left( \frac{\Gamma_{p\rightarrow s}}{\Gamma_{p\rightarrow s}+\Gamma}\right)\left(\frac{\Gamma}{\Gamma + 2\gamma}\right),
$$
where the timing-jitter captured by the first factor acts to suppress the indistinguishability.

%\subsection{Indistinguishability correction}

%Mathematica method with lifetime from g2
To determine the indistinguishability of the emitted single photons we made use of an unbalanced free beam MZI. 
The contrast of the MZI has been ascertained by making use of an narrowband infrared diode laser whose wavelength is similar to the QD. 
The Michelson contrast $C_M$ has been measured by piezo-shifting the delay arm of the MZI to record the maximum and minimum of the 
laser power $I_{\mathrm{max}} = (533.0 \pm 5.3)\,\mu W$ and $I_{\mathrm{min}} = (2.80 \pm 0.03)\,\mu W$. By making use of the expression
\begin{equation}
C_M = 1 - \epsilon = \frac{I_{max}-I_{min}}{I_{max}+I_{min}},
\label{contrast}
\end{equation}
we obtained a Michelson contrast of $C_M = (99.0\,^{+1.0}_{-2.0})\,\%$. Additionally the reflectivity $R$ and transmission $T$ of the 
imperfect 50:50 beamsplitter were determined to be $R = 0.485$ and $T = 0.515$. 
We made use of the correction given by Santori {\it{et al.}}~\cite{Santori2002} to correct the area of the central peak of the 
coincidence histogram for parallel and orthogonal polarization according to the expressions
\begin{equation}
A_{\parallel} \propto (R^{3}T + RT^{3}) \cdot (1 + 2g^{*}) - 2(1 - \epsilon)^{2}\,R^{2}T^{2}\,\nu
\label{A_para}
\end{equation}
\begin{equation}
A_{\perp} \propto (R^{3}T + RT^{3}) \cdot (1 + 2g^{*}),
\label{A_perp}
\end{equation}
where $g^{*}$ expresses the $g^{(2)}(0)$-value given by only taking into account the adjacent peaks at $\pm\,12.2$\,ns.

\end{appendix}

\end{document}